\documentclass[twocolumn,apj,floats,aps,amsmath,amssymb,nofootinbib,numberedappendix,appendixfloats,twocolappendix]{emulateapj}
\usepackage{graphicx}
\usepackage{bm}
\usepackage{hyperref}
\hypersetup{pdfauthor=Hans Baehr}
\hypersetup{backref=true, pagebackref=true, hyperindex=true, breaklinks=true, colorlinks=true, urlcolor=blue}
\hypersetup{citecolor=blue, pagecolor=red, bookmarks=true, bookmarksopen=true}
\usepackage[english]{babel}
\usepackage{tabu}
\usepackage{amsmath}
\usepackage[caption=false, position=t, singlelinecheck=off]{subfig}

\begin{document}

\citestyle{egu}
\bibliographystyle{apj}

\title{The fragmentation criteria in local vertically stratified self-gravitating disk simulations}
\author{Hans Baehr$^{1,*}$}
\author{Hubert Klahr$^1$}
\author{Kaitlin M. Kratter$^2$}

\affil{$^1$Max Planck Institute for Astronomy, K{\"o}nigstuhl 17, D-69117 Heidelberg, Germany}
\affil{$^2$Steward Observatory, University of Arizona, Tucson, AZ 85721, USA}

\email{baehr@mpia.de}
\altaffiltext{*}{Fellow of the International Max Planck Research School for Astronomy and Cosmic Physics at the University of Heidelberg (IMPRS-HD)}

\begin{abstract}
Massive circumstellar disks are prone to gravitational instabilities, which trigger the formation of spiral arms that can fragment into bound clumps under the right conditions. Two dimensional simulations of self-gravitating disks are useful starting points for studying fragmentation, allowing for high-resolution simulations of thin disks. However, convergence issues can arise in 2D from various sources. One of these sources is the 2D approximation of self-gravity, which exaggerates the effect of self-gravity on small scales when the potential is not smoothed to account for the assumed vertical extent of the disk. This effect is enhanced by increased resolution, resulting in fragmentation at longer cooling timescales $\beta$. If true, it suggests that the 3D simulations of disk fragmentation may not have the same convergence problem and could be used to examine the nature of fragmentation without smoothing self-gravity on scales similar to the disk scale height. To that end, we have carried out local 3D self-gravitating disk simulations with simple $\beta$ cooling with fixed background irradiation to determine if 3D is necessary to properly describe disk fragmentation. Above a resolution of $\sim 40$ grid cells per scale height, we find that our simulations converge with respect to the cooling timescale. This result converges in agreement with analytic expectations which place a fragmentation boundary at $\beta_\mathrm{crit} = 3$.
\end{abstract}


\maketitle

\section{Introduction}
\label{sec:intro}

Planet formation in circumstellar disks relies largely on the core accretion to explain most of the planets observed to date. The slow build up of solids to form planetesimals and planetary cores is sufficient to explain most discovered planets within $\sim 10$ AU of their host star \citep{Safronov1969,Lissauer1993,Mordasini2012}. The formation timescale of core accretion increases with distance from the star, making the formation of gas giant planets at wide radii more difficult. Fragmentation through gravitational instability (GI) is a model which forms stellar companions quickly and typically only works in the cool outer reaches of the circumstellar disk \citep{Kuiper1951,Boss1997}. When GI fails to produce a companion star, it may result in the formation of bound companions with masses on the order of several Jupiter masses.

Gravitational instability becomes relevant when a circumstellar disk has enough gas mass to form local overdensities that may collapse under their own self-gravity. Collapse only proceeds when the disk cools efficiently enough to suppress thermal pressure support  \citep{Gammie2001,Durisen2007}. These two conditions are measured by the Toomre parameter $Q$ and a simple cooling timescale $t_{\textnormal{c}}$, respectively. The Toomre parameter quantifies the balance of stabilizing effects of rotation and pressure versus the destabilization generated by gas overdensities and is defined in a linear 2D approximation by \citet{Toomre1964} as:
\begin{equation} \label{eq:toomre}
Q = \frac{c_{\textnormal{s}}\Omega}{\pi G\Sigma},
\end{equation}
where $\Omega = \sqrt{GM / R^{3}}$ is the orbital Keplerian frequency of the disk, stabilizing the disk to density perturbations at larger wavelengths, and $c_{\textnormal{s}}$ is the local speed of sound, which stabilizes the disk to density perturbations at smaller wavelengths \citep{Durisen2007}.  Theoretical models of protoplanetary disks often assume a thin disk, so quantities are vertically integrated, i.e. surface density $\Sigma \approx \rho H$, where $H$ is the scale height of the disk. A critical balance between stability and instability is defined for $Q=1$. Areas with $Q \lesssim 1$ are unstable to axisymmetric density perturbations in the linear regime, which may contract and fragment following the nonlinear growth of the perturbations, while a region with $Q \gtrsim 1$ will remain stable to collapse \citep{Toomre1964}. When marginally stable, a disk tends to stay close to $Q \gtrsim 1$ . Falling below $Q = 1$ generates strong shock heating as structures begin gravitationally contracting, increasing temperatures \citep{Balbus1999}. The balance of shock heating and cooling results in a gravitoturbulent self-regulating disk, subject to instabilities, but unable to collapse into fragments without an additional constraint.

With a disk balancing on the edge of instability, the cooling timescale is second requirement for fragmentation. The cooling timescale as used here, defines the rate at which the disk loses all of its internal energy. Thus, if the disk can cool efficiently, comparable to the dynamical timescale $\Omega^{-1}$, thermal pressure support is removed quickly enough to allow for collapse by self-gravity. When cooling is too slow, thermal saturation will provide pressure strong enough to push back against collapse. Therefore, using a simple prescription for the cooling timescale
\begin{equation} \label{eq:cool}
t_{c} = \beta\Omega^{-1},
\end{equation}
where $\beta$ is constant of around unity for efficient cooling and leads to fragmentation with a critical boundary at $\beta_{\mathrm{crit}} = 3$ \citep{Gammie2001}. This critical cooling threshold between the gravitoturbulent state and fragmentation is dependent on adiabatic index $\gamma$ which will push the critical $\beta$ value up slightly for lower values of $\gamma$ \citep{Rice2005}.

Both SPH and grid simulations have failed to converge to a critical cooling parameter \citep{Meru2011b,Paardekooper2012,Lodato2011}. Potential resolutions to this issue have varied from changes to the implementation of viscosity \citep{Rice2014,Klee2017}, to varied cooling methods \citep{Rice2012,Baehr2015}, but one likely reason for the problem lies in the smoothing of the self-gravity potential \citep{Young2015}. Without smoothing self-gravity to approximately the scale height of the disk \citep{Hure2009,Muller2012}, self-gravity is exaggerated in 2D, the effect being more pronounced with increasing resolution, resulting in easier runaway collapse with each increase in resolution. Thus, 3D models are needed to clarify the critical cooling boundary for disk fragmentation including what value the boundary takes and whether or not it converges in 3D with a simple cooling law including irradiation. It is still prudent however, to monitor the effect of various dissipation strengths on the behavior of fragmentation.

Global 3D studies of disk fragmentation are common \citep{Mejia2005,Boley2006,Mayer2007,Michael2012,Vorobyov2013,Lichtenberg2015} but because one simulates the entire disk, it is difficult to resolve the small scales of clump/fragment formation. Using local 3D high-resolution simulations, we can instead focus on a small area of the disk and its interaction with embedded objects. Recently, an increasing number of \emph{local} 3D simulations have been undertaken, but do not focus on the formation of fragments \citep{Hirose2017,Riols}. Their focus has been on properties of the gravitoturbulent state, such as the nature of turbulence and non-axisymmetric structure \citep{Mamatsashvili2010,Shi2014} and dust concentration \citep{Shi2013}. \citet{Shi2014} and \citet{Hirose2017} came the closest to establishing a link to the critical cooling threshold of \citet{Gammie2001} using 3D local simulations, but stopped short of modeling clump collapse or conducting simulations of the fragmentation threshold over multiple resolutions. Studies of similar 3D physics at galactic scales were conducted by \citet{Kim2002}, which also found a critical Toomre parameter $Q \approx 0.7$ but did not include thermodynamics or investigate the effect of resolution on the fragmentation behavior.

In this paper we investigate the fragmentation boundary in 3D local disks over several resolutions to study the convergence behavior of self-gravitating disks. In light of recent unsuccessful \citep{Baehr2015} and successful \citep{Young2015,Deng} attempts at convergence in 2D self-gravitating disks, we seek to determine if 3D modeling of fragmentation with simple cooling including a fixed floor temperature is closer to a realistic treatment the self-gravity of the problem and results in convergence. This paper will proceed with a brief overview of the relevant hydrodynamic equations in Section \ref{sec:theory} followed by a short look at the numerical setup in Section \ref{sec:numerical}. Finally, we look at the results of our vertically stratified simulations in Section \ref{sec:results} and discuss the implications and consequences on disk fragmentation and the simulation thereof in Section \ref{sec:discussion}.

\section{Theory}
\label{sec:theory}

We consider a non-magnetized, self-gravitating disk in the shearing box approximation. Accordingly, the disk is modeled locally on a small radial-azimuthal patch of the disk, transformed from global cylindrical coordinates $(r,\theta,\phi)$ to local Cartesian coordinates $(x,y,z)$ co-rotating with the disk \citep{Goldreich1965}. These assumptions allow for the modeling of the local properties of the disk while following the evolution of fragments that form when using periodic boundary conditions. Using this model, the relevant conservation equations are shown in Equations \eqref{eq:finalmassconserve}-\eqref{eq:finalenergyconserve}, with terms for the Coriolis effect $2\Omega\times\mathbf{u}$ and centrifugal force $q\Omega v_{x}\mathbf{\hat{y}}$ as well as heating $H$ and cooling terms $C$ in the energy equation.
\begin{align}
\frac{\partial {\rho}}{\partial t} &- q\Omega x\frac{\partial {\rho}}{\partial y} + \nabla\cdot(\rho\mathbf{u}) = f_{D}(\rho) \label{eq:finalmassconserve} \\
\frac{\partial \mathbf{u}}{\partial t} &- q\Omega x\frac{\partial \mathbf{u}}{\partial y} + \mathbf{u}\cdot\nabla\mathbf{u} = \nonumber \\ 
& -\frac{\nabla P}{\rho} + q\Omega v_{x}\mathbf{\hat{y}} - 2\Omega\times\mathbf{u} - \nabla\Phi + f_{\nu}(\mathbf{u}) \label{eq:finalmomconserve} \\
\frac{\partial s}{\partial t} &-q\Omega x\frac{\partial s}{\partial y} + (\mathbf{u} \cdot \nabla)s = \nonumber \\
& \frac{1}{\rho T} \left( 2\rho\nu\mathbf{S}^{2} - \Lambda + f_{\chi}(s) \right) \label{eq:finalenergyconserve}
\end{align}

Here, $\mathbf{u} = (v_{\textnormal{x}},v_{\textnormal{y}}+q\Omega x,v_{\textnormal{z}})^{T}$ is the flow plus shear velocity in the local box, $\rho$ is the gas density, and $s$ is the gas entropy. Viscous heat is generated by $\mathcal{H} = 2\rho\nu\mathbf{S}^{2}$, with rate-of-strain tensor $\mathbf{S}$, and radiated away by a simple $\beta$-cooling prescription $\Lambda$, described in Section \ref{subsec:cool}.

Self-gravity is modeled by solving the Poisson equation
\begin{equation} \label{eq:poisson}
\nabla^{2}\Phi=4\pi G\rho,
\end{equation}
in Fourier space by transforming the surface density to find the potential at the scale of wavenumber $k$ and transforming the solution back into real space. The solution to the Poisson equation in Fourier space is
\begin{equation} \label{eq:gravpotential}
\Phi(\mathbf{k}, t) = -\frac{2\pi G\rho(\mathbf{k}, t)}{\mathbf{k}^2}.
\end{equation}
In 2D simulations it is useful to smooth the self-gravity potential at small scales. When using a Fourier method, this means limiting the wavenumbers included in the calculation of the potential to exclude those corresponding to gravitational attraction at scales smaller than a pressure scale height. This is used in 2D to ascribe a thickness to the disk, however since our simulations include a vertically stratified density distribution, there should be no need for smoothing in these simulations.

Finally, we use an ideal equation of state, with internal energy $\epsilon$, and specific heat ratio $\gamma$
\begin{equation} \label{eq:eos}
P = (\gamma - 1)\rho\epsilon.
\end{equation}
We use a ratio of $\gamma = 5/3$, which corresponds to a 2D value of $\gamma = 1.8$ \citep{Gammie2001}.

\begin{table}[t]
\caption{\textnormal{3D Simulations at various resolutions $N^3$, initial Toomre criteria $Q_0$, and cooling times $\beta$.}}
\centering
\begin{tabular}{l*{6}{c}r}
Simulation  & Grid Cells ($N^3$) &$Q_0$ & $\beta$ & Fragmentation \\
\hline
Q256t2            & $256^3$  & 1      & 2   & No \\
Q256t10           & $256^3$  & 1      & 10  & No \\
Q512t2            & $512^3$  & 1      & 2   & No \\
Q512t10           & $512^3$  & 1      & 10  & No \\\noalign{\vskip 1.5mm}
G512tp5           & $256^3$  & 0.676  & 0.5 & No \\
G256t1            & $256^3$  & 0.676  & 1   & No \\
G256t2            & $256^3$  & 0.676  & 2   & No \\
G256t10           & $256^3$  & 0.676  & 10  & No \\\noalign{\vskip 1.5mm}
G512t2            & $512^3$  & 0.676  & 2   & Yes \\
G512t5            & $512^3$  & 0.676  & 5   & No \\
G512t10           & $512^3$  & 0.676  & 10  & No \\
G512t20           & $512^3$  & 0.676  & 20  & No \\\noalign{\vskip 1.5mm}
G1024t2           & $1024^3$ & 0.676  & 2   & Yes \\
G1024t5           & $1024^3$ & 0.676  & 5   & No \\
G1024t10          & $1024^3$ & 0.676  & 10  & No \\
G1024t20          & $1024^3$ & 0.676  & 20  & No \\
\end{tabular}
\label{tab:sims}
\end{table}

  \subsection{Viscous Stress}
  \label{subsec:alpha}

Accretion driven by turbulence during the early self-gravitating phase of a disk is expected to be strong, and according to the $\alpha$-formalism \citep{Shakura1973}, typically approaches unity $\alpha \approx 0.1-1$, a few orders of magnitude higher than that of magnetorotational instability \citep{Balbus1991}. To measure the $\alpha$-stress in our simulations, we will need to consider the contribution of the gravitational and Reynolds stresses:
\begin{equation} \label{eq:alpha}
\alpha = -\left(\frac{d\ln\Omega}{d\ln r}\right)^{-1}\frac{\langle G_{xy}\rangle + \langle R_{xy}\rangle}{\rho c_{s}^{2}},
\end{equation}
where
\begin{equation}
\langle G_{xy}\rangle = \int_{-\infty}^{\infty} \frac{g_{x} g_{y}}{4 \pi G} dz,
\end{equation}
and
\begin{equation}
\langle R_{xy}\rangle = \langle \rho u_{x} u_{y} \rangle.
\end{equation}

An analytic expectation of $\alpha$ in a gravitoturbulent thin disk is given by \citet{Gammie2001}
\begin{equation} \label{eq:alphapara}
\alpha = \frac{4}{9}\frac{1}{\gamma(\gamma-1) t_{c}\Omega}.
\end{equation}
\citet{Shi2014} compare the standard formulation of $\alpha$ (Equation \eqref{eq:alpha}) to one based on density-weighted values of the pressure and stress and find that while the density-weighted $\alpha'$ is a factor of $1.25$-$2$ higher, the usual formulation more closely follows the theoretical expectation \eqref{eq:alphapara}. From Equations \eqref{eq:alpha} and \eqref{eq:alphapara}, one can compare theoretical expectations of viscous stresses in the gravitoturbulent state with the stresses generated by our simulations. It is important to note that as a local description of viscous stress, $\alpha$ does not necessarily model gravitationally unstable disks, as gravity inherently acts on non-local scales \citep{Balbus1999}. 

  \subsection{Thermodynamics}
  \label{subsec:cool} 

Proper modeling of disk cooling is complex even without using sophisticated ray-tracing and flux-limited diffusion schemes, depending significantly on the local temperature and density of the disk from midplane to atmosphere \citep{Hubeny1990,Rafikov2005}
\begin{equation} \label{eq:bettercooling}
t_{\textnormal{c}} = \frac{4}{9 \gamma (\gamma - 1)}\frac{\Sigma c_{\textnormal{s}}^{2}}{\sigma (T^4 - T_{\textnormal{irr}}^{4})} f(\tau),
\end{equation}
where $\sigma$ is the Stefan-Boltzmann constant, $f(\tau)=\tau + 1/\tau$ for optical depth $\tau$, $T_{\textnormal{irr}}$ is the reference temperature and adiabatic index $\gamma$.

Contraction to a fragment requires that thermal pressure support does not counter the self-gravity of overdense perturbations. Heat generated through viscous heating and shock dissipation should be released on a cooling timescale short enough to keep pressure support from building up, or shorter than a few dynamical timescales (see Equation \eqref{eq:cool}). Cooling with a $\beta$-prescription is implemented in Pencil using a Newtonian cooling scheme
\begin{equation} \label{eq:coolinglaw2}
\Lambda = \frac{\rho (c_{\textnormal{s}}^{2} - c_{\textnormal{s,irr}}^{2})}{(\gamma -1) t_{\textnormal{c}}},
\end{equation}
where $c_{\textnormal{s,irr}}^{2}$ is proportional to a non-zero background temperature, corresponding to some constant irradiation, which as demonstrated by \citet{Lin2016} adds additional stability against small scale density perturbations, and $t_{\textnormal{c}}$ is a $\beta$-cooling timescale \eqref{eq:cool}. The sound speed relates to entropy as 
\begin{equation} \label{eq:entropy}
c_{\textnormal{s}}^2 = \gamma \frac{P}{\rho} = c_{\textnormal{s,0}}^2 \textnormal{exp}(\gamma s/c_{\textnormal{p}} + (\gamma - 1) \ln (\rho / \rho_0)), \end{equation}
where $c_{\mathrm{p}} = 1$, initial sound speed $c_{\textnormal{s,0}}=\pi$, density $\rho_0=0.293$ and temperatures are derived from the sound speed using $T=c_{\mathrm{s}}^2/c_{\mathrm{p}} (\gamma - 1)$. This is a simple way to model disk cooling which is not as computationally expensive as more realistic methods, particularly at the resolutions used here.  

The cooling described in Equation \eqref{eq:coolinglaw2}, is balanced by two sources of heat generation: viscous heating and shock dissipation. Viscous heating is calculated as 
\begin{equation} \label{eq:viscousheat}
\mathcal{H} = 2\rho\nu\mathbf{S}^{2},
\end{equation}
with kinematic viscosity $\nu = 2.5 H^6 \Omega$ for all simulations, as seen in Equation \eqref{eq:finalenergyconserve} but ultimately the heating of the simulation will be dominated by the dissipation of shocks produced by the collapse of self-gravitating filaments and clumps. The heat generated by the dissipation of shocks is determined by the strength of the constant of hyperdissipation so in addition to our standard collection of simulations spanning resolution and cooling timescale, we include a small sample of simulations with various dissipation values to note the effect on fragmentation.

In practice, this may be overestimating the heating resulting from strong shocks generated by self-gravity. It is well understood that in outer disk regions, disk heating is dominated by stellar irradiation instead of viscous accretion \citep{DAlessio1999a}. Additionally, recent work by \citet{Rafikov2016} suggests that shocks are not a dominant source of heat in passive disks such as those considered here.
%
%
%

  \subsection{Vertical Stratification}
  \label{subsec:stratification}

A proper implementation of a 3D shearing box simulation requires vertical stratification of the gas density. If this is not correctly initialized, the simulation will adjust on its own, causing undesired rapid heating and cooling as the disk swells up or contracts. Thus we assume the disk begins in hydrostatic equilibrium with self-gravity
\begin{equation} \label{eq:hydrostat}
\frac{1}{\rho} \frac{dP}{dz} = -\Omega^2 z - 4\pi G \int_0^z \rho(z^\prime) dz^\prime.
\end{equation}
The initial vertical density distribution is determined by rewriting the above as  a dimensionless, differential equation, following \citet{Shi2014}
\begin{equation} \label{eq:hydrostatode}
\frac{d^2\tilde{\rho}}{d\tilde{z}^2} + \frac{\gamma - 2}{\tilde{\rho}} \left(\frac{d\tilde{\rho}}{d\tilde{z}}\right)^2 + Q_{0}^2 \tilde{\rho}^{2-\gamma} + \frac{2}{h}\tilde{\rho}^{3-\gamma} = 0.
\end{equation}
We use an isothermal approximation to the polytropic equation of state for which Equation \eqref{eq:hydrostatode} was derived ($\gamma = 1, K = c_{\textnormal{s}}^2$) and assume $\Sigma_0=2H\rho$. Solving for vertical hydrostatic equilibrium including self-gravity we find
\begin{equation} \label{eq:hydrostat2}
\frac{d^2\tilde{\rho}}{d\tilde{z}^2} - \frac{1}{\tilde{\rho}} \left(\frac{d\tilde{\rho}}{d\tilde{z}}\right)^2 + Q_{0}^2 \tilde{\rho} + \frac{2}{h}\tilde{\rho}^2 = 0,
\end{equation}
where $\tilde{\rho}=\rho/\rho_0$, $\tilde{z}=z/(c_{\textnormal{s,0}}^{2}/\pi G\Sigma_0)$, $Q_0=c_{\textnormal{s,0}}\Omega/\pi G\Sigma_0$ and $h=H/(c_{\textnormal{s,0}}^{2}/\pi G\Sigma_0)$. The approximation of the vertically stratified initial density creates some small oscillations in as the simulation equilibriates to the true solution. This results in a small flux of mass away from the regions between $1<|z|<2$ and towards the outer disk atmosphere, but these result in density deviations on the order of $10^{-4}$ and do not have a significant impact on the outcome of the simulation. Due to the large empty regions above and below the disk midplane, it is necessary to impose a minimum density value to avoid extreme density differences which would slow down the code immensely.

\begin{figure*}[t]
\centering
\includegraphics[width=0.5\textwidth]{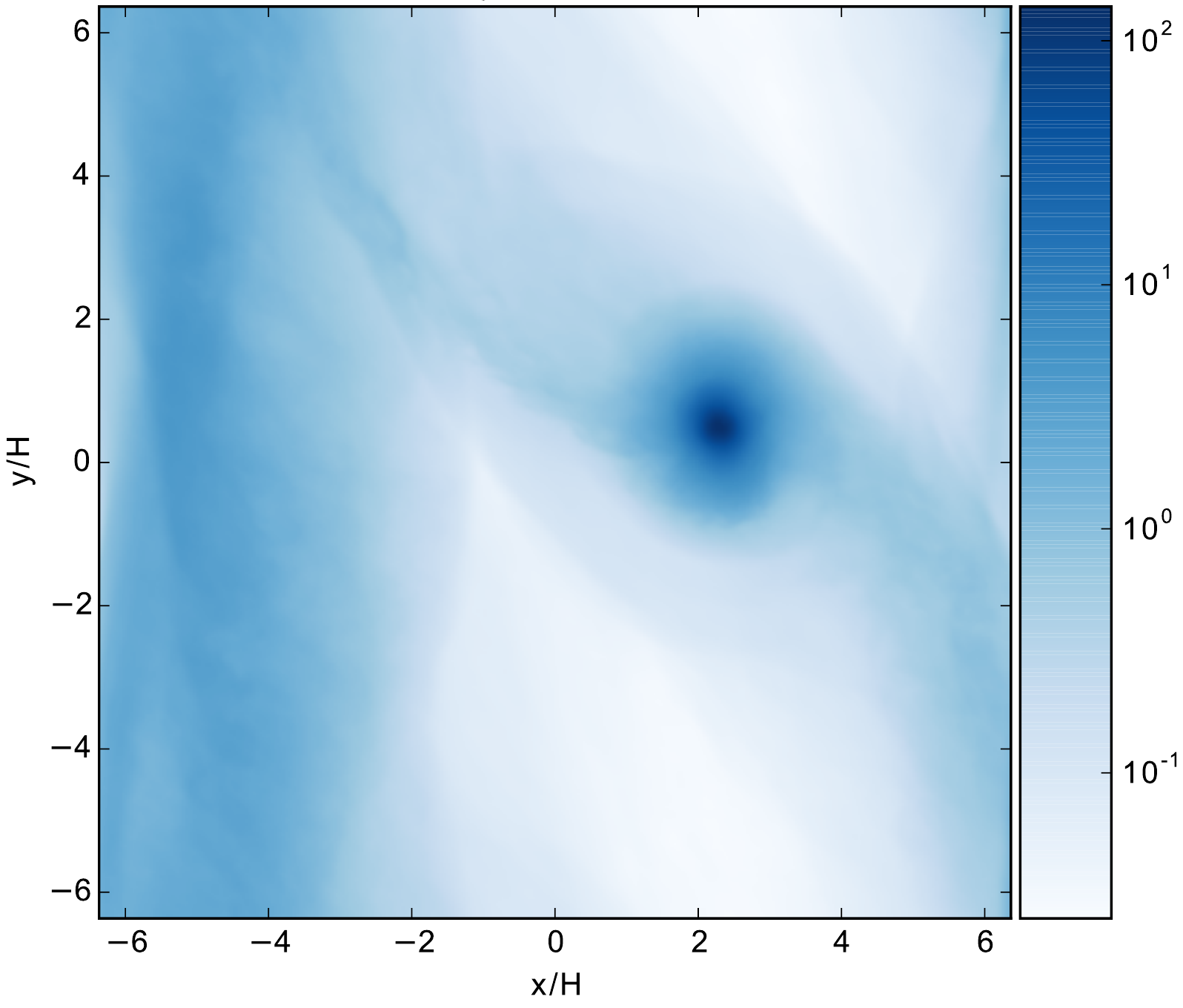}%
\includegraphics[width=0.5\textwidth]{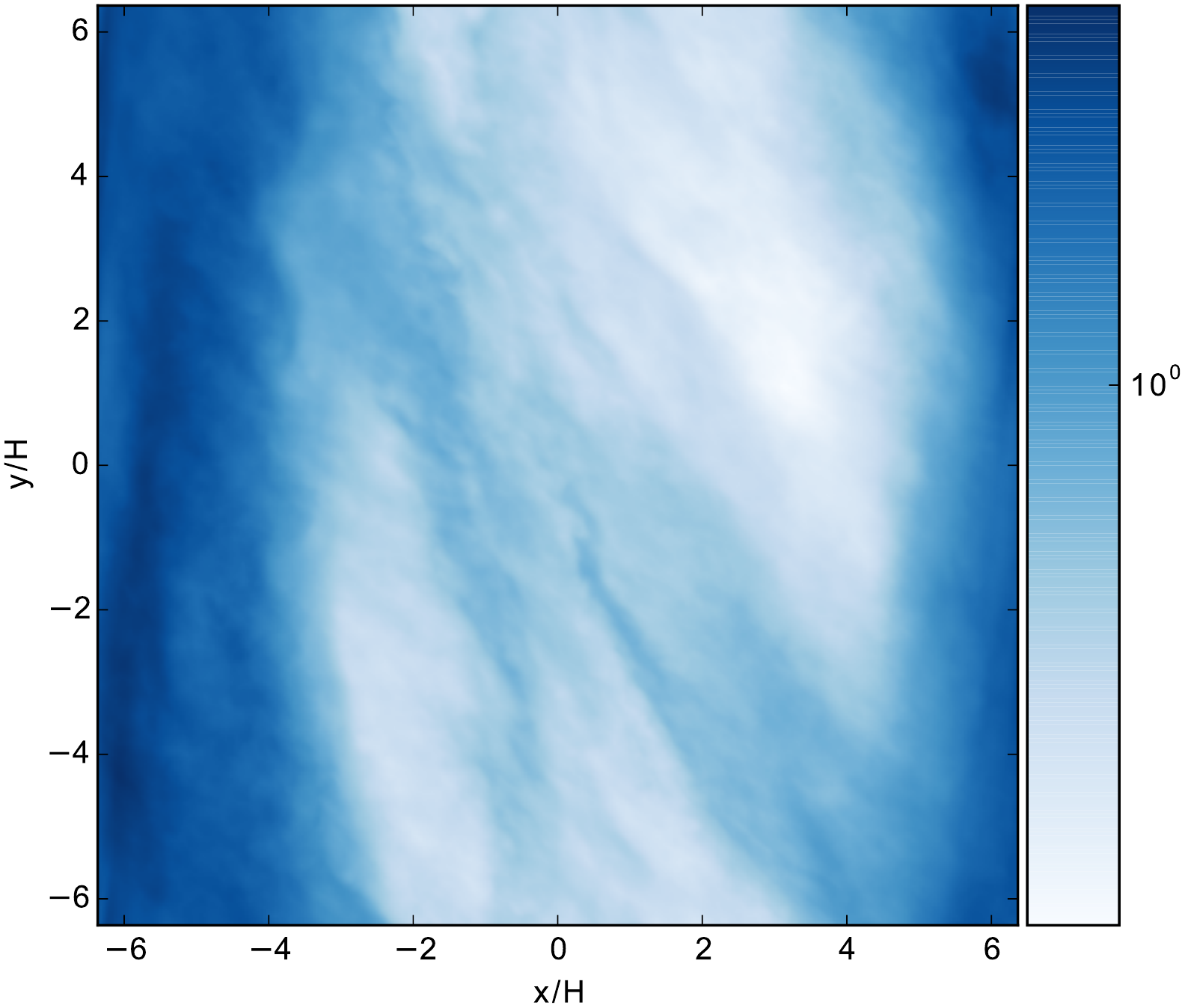}%
\caption{Surface density maps of a pair of our medium resolution ($N^3=512^3$) runs. On the left is a simulation with short $\beta=2$ cooling, illustrating the clear formation of a fragment. On the right, a simulation with $\beta=10$ cooling showing steady gravitoturbulence.}
\label{fig:fragment}
\end{figure*}

The initial temperature distribution is not stratified but instead isothermal throughout the entire vertical extent, equal to the background irradiation level. Since distant regions of the disk will be dominated by disk irradiation rather than accretion heating this is a reasonable assumption. As the disk settles to hydrostatic equilibrium, shocks are generated from self-gravitational collapse and generate significant heat, but cooling returns the vertical temperature profile to its original flat distribution.

  \subsection{Comparison to the Thin Disk Approximation}
  \label{subsec:thindisk}

Introducing vertical structure affects the stability and dynamics of a simulation, and should be accounted for in comparison to a thin disk approximation. Mass density should be stratified so that it is as close to hydrostatic equilibrium as possible, otherwise the simulation will adjust itself and if the difference is significant, the simulation will be adversely affected. In a simulation on the border between fragmentation and stability, too much heating initial heating may suppress fragmentation and too much cooling may result in premature fragmentation.

The stratification of mass density also results in the weakening of self-gravity, meaning a simulation established with $Q_0=1$ will not become unstable. In the thin disk approximation, the gas density is uniformly distributed over two scale heights, but in our 3D simulations, it is instead concentrated in at the midplane but distributed over twelve vertical scale heights. With more mass further away from the midplane, a disk needs to be more massive to become unstable. Theoretical expectations indicate a critical Toomre criterion closer to $Q_0 \approx 0.7$ \citep{Behrendt2015,Kratter2016} and our simulations reflect that.

\section{Numerical Methods}
\label{sec:numerical}

All simulations were conducted using the Pencil code\footnote{http://pencil-code.nordita.org/} \citep{Brandenburg2001}, which is open source and no additional modifications were made in the implementation of these runs.

  \subsection{Artificial Viscosity}
  \label{subsec:artvisc}

An important question regarding the stability of self-gravity is the role of artificial viscosity on the onset of fragmentation. We use a hyperdiffusion method, implemented in Pencil according to Appendix B of \citet{Yang2012}
\begin{equation} \label{eq:hyperdiff}
f_{D}(\Sigma) = \zeta_{D}(\nabla^{6}\Sigma),
\end{equation}
which smooths perturbations on small scales without affecting the power on larger scales. It has been suggested that such viscosities are the reason for the non-convergence of the 2D simulations \citep{Rice2014}. The choice of a stronger dissipation scheme, such as hyperviscosity, is justified because of the effects of numerical oversteepening \citep{Klee2017}, which may lead to the overestimation of overdense perturbations and contribute to non-convergent behavior.

  \subsection{Boundary Conditions}
  \label{subsec:bound}

The shearing box is set up with periodic $y$ and $z$ boundaries, corresponding to the azimuthal and radial physical extents, respectively. This means that all quantities which exit one boundary are reproduced at the other side of the box, including the self-gravitational potential in the vertical direction.
\begin{align} 
& f(x, y, z, t) = f(x, y + L, z, t) \label{eq:ybounds} \\
& f(x, y, z, t) = f(x, y, z + L, t) \label{eq:zbounds}
\end{align}
The boundaries $x = 0$ and $x = L$ require a different boundary condition on account of the shear velocity $u_{\textnormal{y}} = v_{\textnormal{y}} + \frac{3}{2}\Omega x$ \citep{Hawley1995}.
\begin{equation} \label{eq:xbounds}
f(x, y, z, t) = f(x + L, y - \frac{3}{2}\Omega Lt, z, t)
\end{equation}
When calculating this over a shear periodic x-boundary, the displacement due to the shear is taken into account by shifting the entire y-direction to make the x-direction periodic before proceeding with the transform in the x-direction. After the calculation of the potential in Fourier space the process is reversed to get back to real space.

  \subsection{Initial Conditions}
  \label{subsec:init}

Initial values of density and sound speed are set by establishing an overall Toomre $Q$ (Equation \eqref{eq:toomre}) which is borderline unstable. To this end, we use $\Omega=G=1$ and must consider how we calculate the sound speed and surface density. By setting the sound speed to $c_{\textnormal{s,0}} = \pi$ we are left with the surface density $\Sigma$ which while normally one would ensure $\Sigma_0 = \int{\rho_{0} (z) dz} = 1$ to achieve $Q_0 = 1$, we find it necessary to also consider the case where $Q_0 \approx 0.676$ and initialize the surface density accordingly. To account for the gravitational field of the central star we include a sinusoidal acceleration profile, such that mass settles towards the midplane but acceleration approaches $0$ at the vertical boundaries. Such a profile is not as realistic as a linear profile, but it improves stability. 

A reference temperature proportional to $c_{\textnormal{s,irr}}^2= c_{\textnormal{s,0}}^2 = \pi^2$ is used in Equation \eqref{eq:coolinglaw2} since the sound speed is initialized as $c_{\textnormal{s,0}} = \pi$. The shearing box needs to be large enough to contain a critical Toomre wavelength $\lambda = 2\pi / k_{\mathrm{cr}} = 2\pi H$, while also resolving this critical wavelength with at least 4 grid cells and avoid artificial fragmentation \citep{Nelson2006}. We meet these conditions using $L_{\textnormal{x}}=L_{\textnormal{y}}=L_{\textnormal{z}}=(40 /\pi)H\approx 12.7H$; see Figure \ref{fig:fragment} for an example of the box scale. These boxes are smaller in radial and azimuthal extent than other comparable simulations \citep{Hirose2017,Riols} allowing for higher resolution per scale height comparatively.

  \subsection{Resolution Requirements}
  \label{subsec:resolution}
\begin{figure*}[t]
\centering
\includegraphics[width=0.5\textwidth]{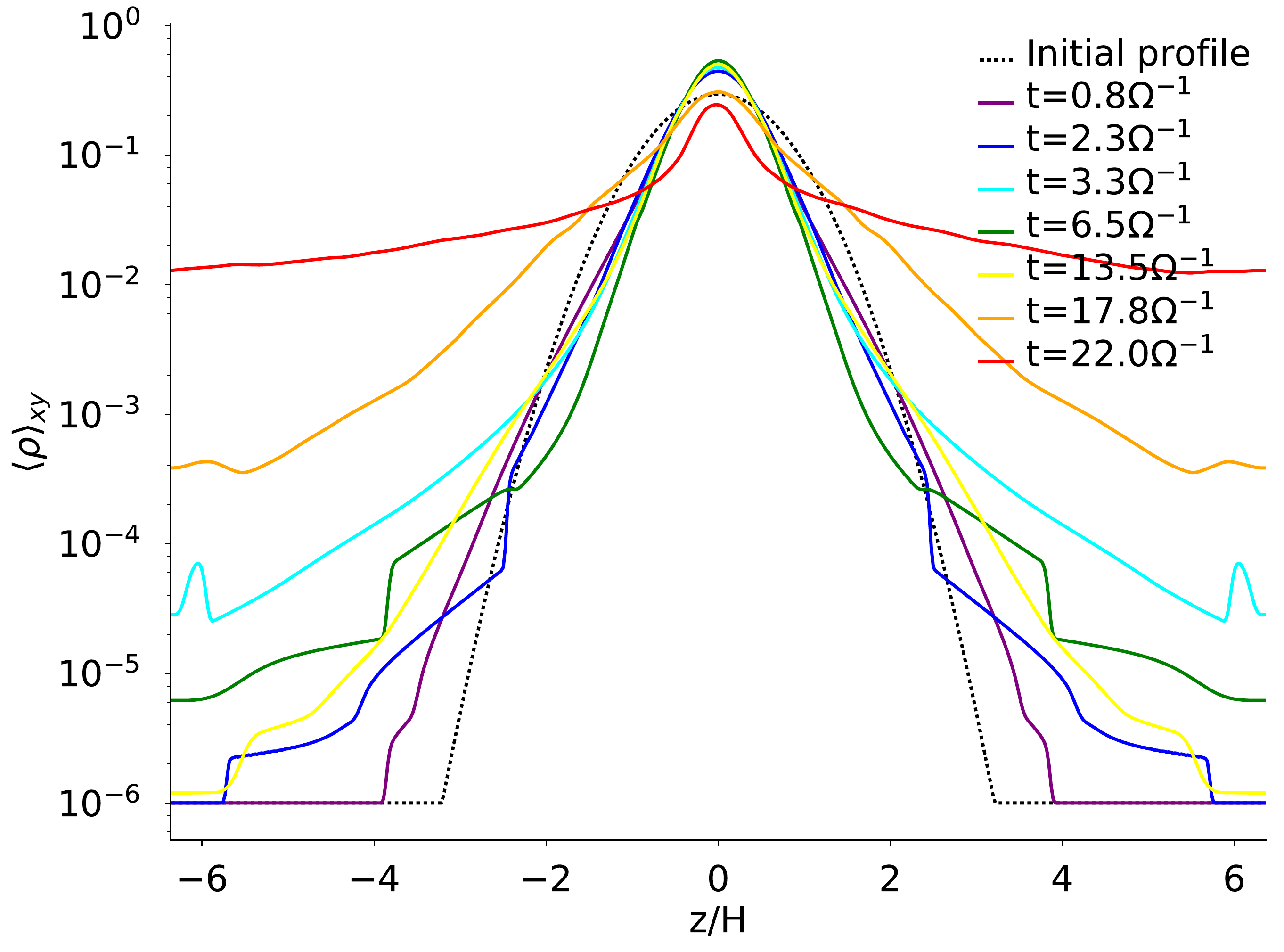}%
\includegraphics[width=0.5\textwidth]{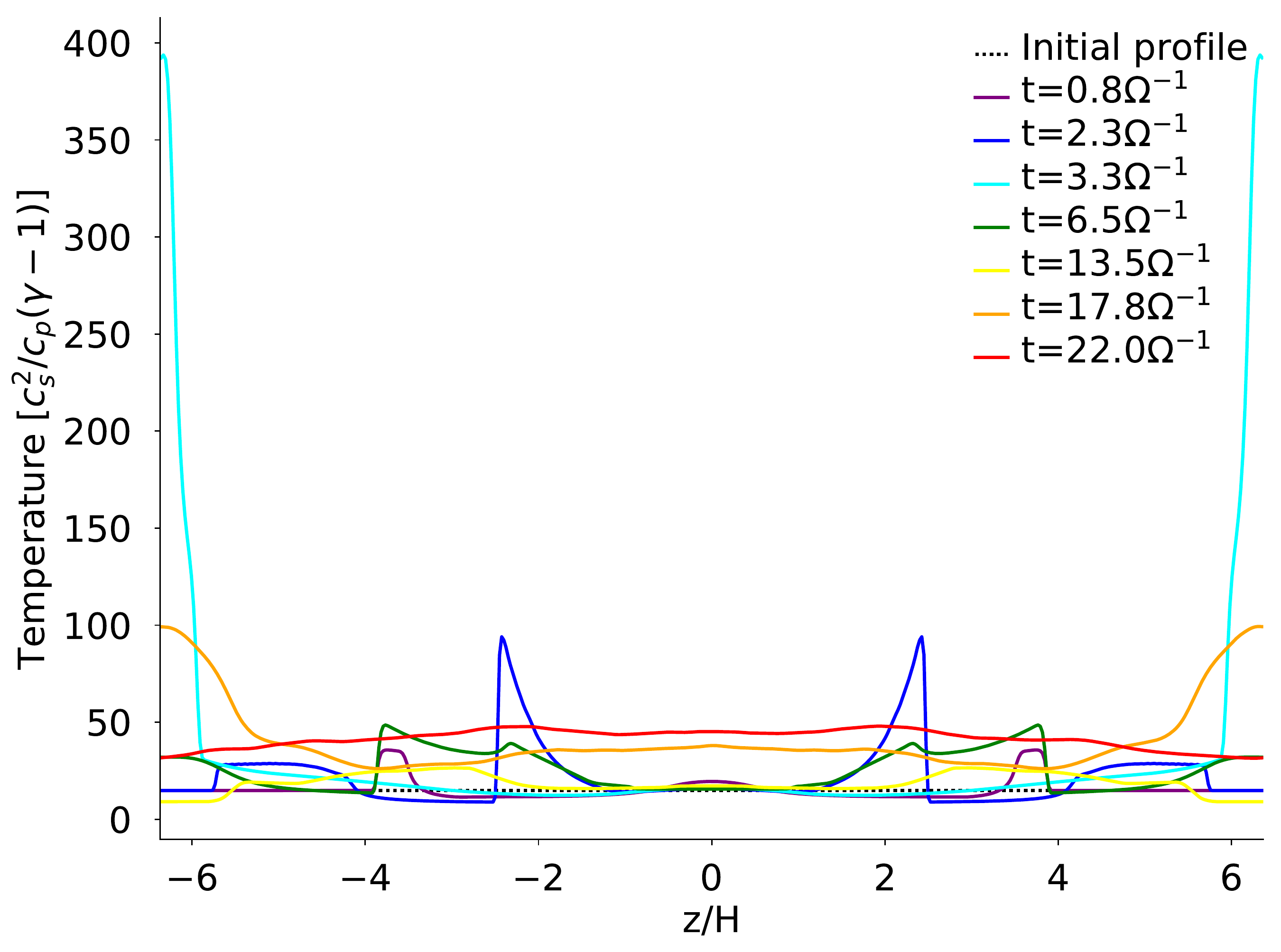}%
\caption{Evolution of the average vertical density (\textit{left}) and average vertical temperature (\textit{right}) profiles in the simulation \textit{G512t2}. Colored lines represent the vertical averages at different times.}
\label{fig:profileslQ}
\end{figure*}
Artificial fragmentation due to poor resolution is a significant concern with self-gravitating simulations, as insufficient resolution can lead to non-negligible errors which form spurious fragments \citep{Truelove1997}. \citet{Nelson2006} found that for gravitationally unstable disk simulations, the critical Toomre wavelength needs to be resolved by at least 4 resolution elements to avoid such fragmentation, described by a maximum Toomre number of 
\begin{equation} \label{eq:toomrenumber}
T = \frac{\delta x}{\lambda_{T}} = 0.25.
\end{equation}
Our simulations contain almost exactly one critical Toomre wavelength and with a resolution of $N^3=256^3$ cells in each direction, our minimum resolution easily meets this resolution criterion with $T=0.008$, decreasing by a factor of one half when increasing the resolution to $N^3=512^3$ and $N^3=1024^3$. The simulations of \citet{Boss2017} only consider fragments at or near this resolution criterion, and are thus potentially susceptible to fragmentation due to numerical errors. Additionally, the Toomre length only roughly describes the feeding zone of an unstable clump which may or may not fragment further \citep{Kratter2011}, and is not necessarily a length indicative of a fragment.
\begin{figure}
\centering
\includegraphics[width=0.5\textwidth]{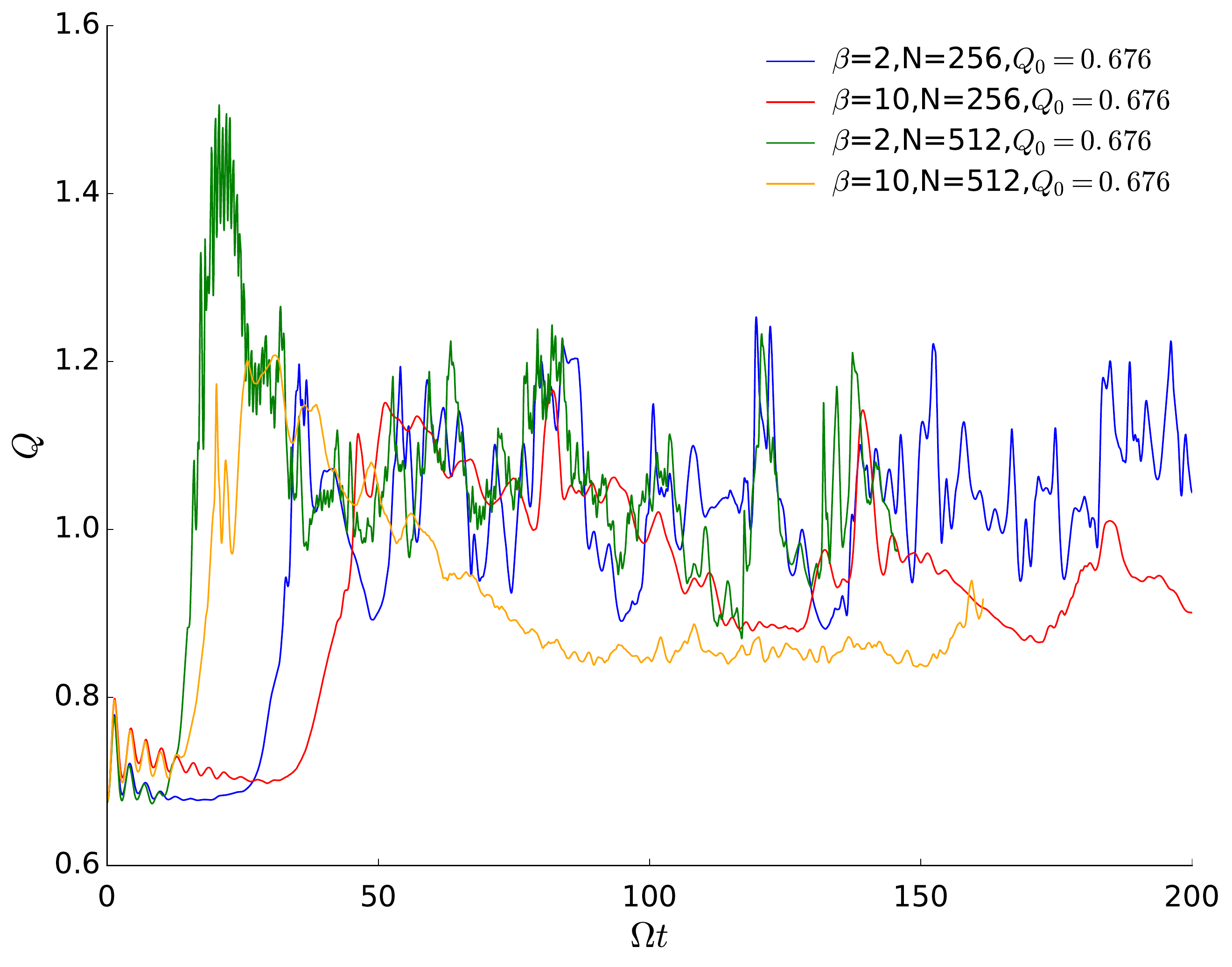}
\caption{Box average Toomre $Q$ some of our low and medium resolution runs. $Q$ was calculated using the density weighted sound speed and the box-averaged sound speed. The single fragmenting case (in green) distinguishes itself from the other gravitoturbulent runs, stabilizing the surrounding disk as the fragment collapses.}
\label{fig:toomre}
\end{figure}
\begin{figure*}[t]
\centering
\includegraphics[width=0.5\textwidth]{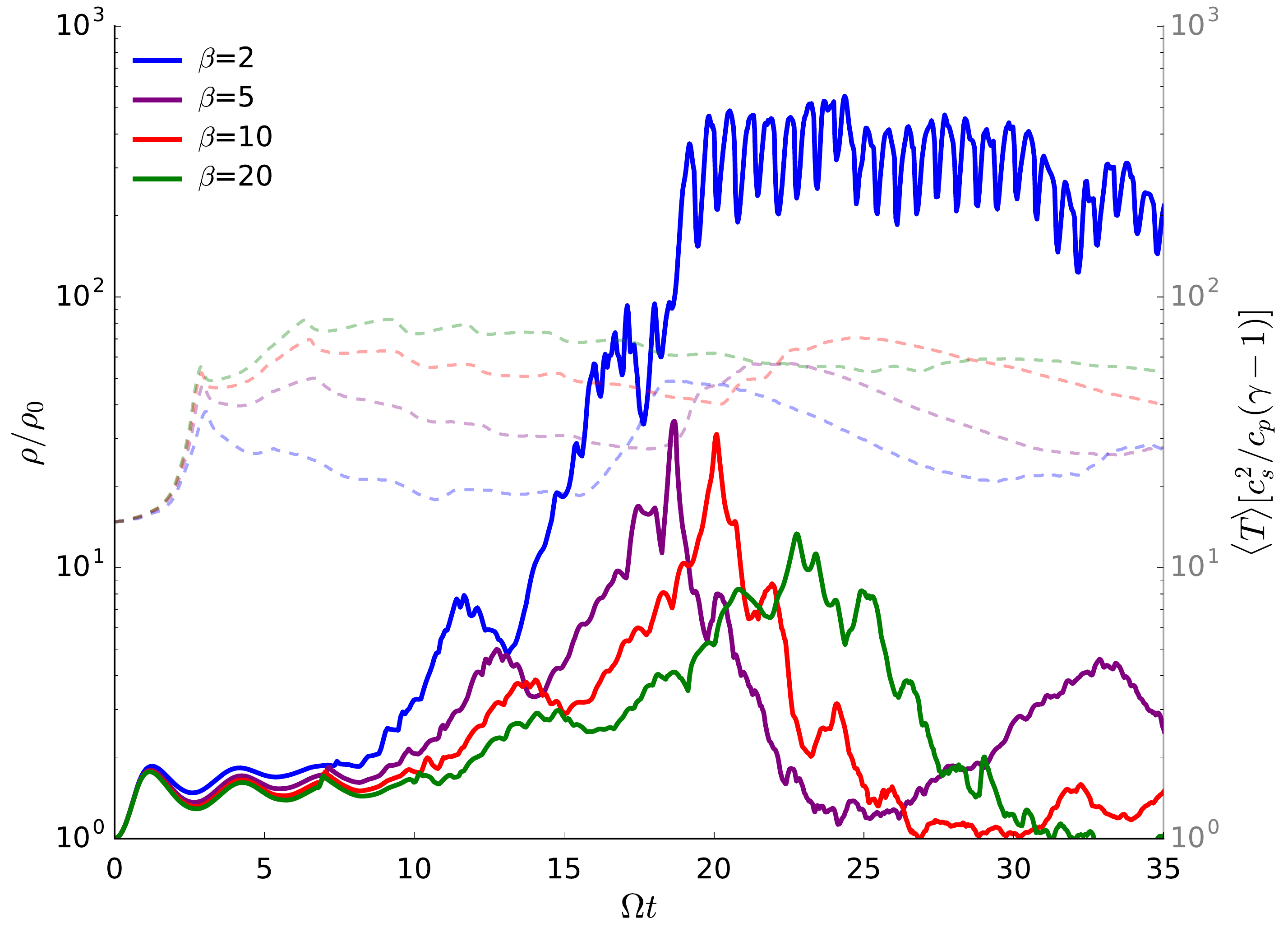}%
\includegraphics[width=0.5\textwidth]{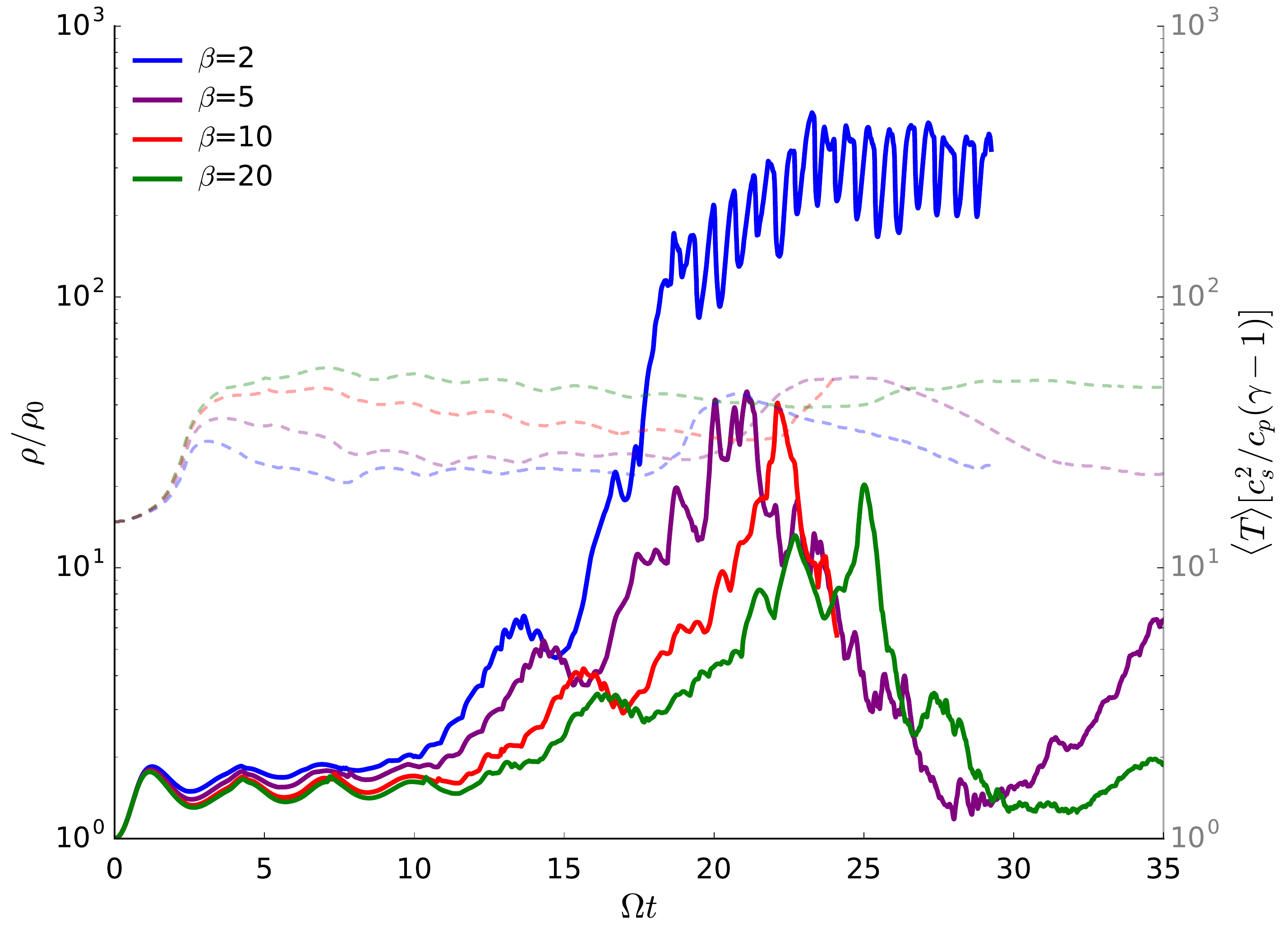}%
\caption{\textit{Left}: The maximum surface densities of four simulations carried out at resolution $N^3=512^3$ and $Q_0 = 0.676$. Fragmentation outcomes are dependent on the cooling timescale, consistent with a $\beta_{\textnormal{crit}} \sim 3$. \textit{Right}: The maximum surface densities of two simulations carried out at resolution $N^3=1024^3$ and $Q_0 = 0.676$. Fragmentation outcomes are based on the cooling timescale, consistent with a $\beta_{\textnormal{crit}} \sim 3$. Fainter solid lines represent the box-averaged temperatures for each simulation.}
\label{fig:maxdensitylQ}
\end{figure*}
\begin{figure}[t]
\centering
\includegraphics[width=0.5\textwidth]{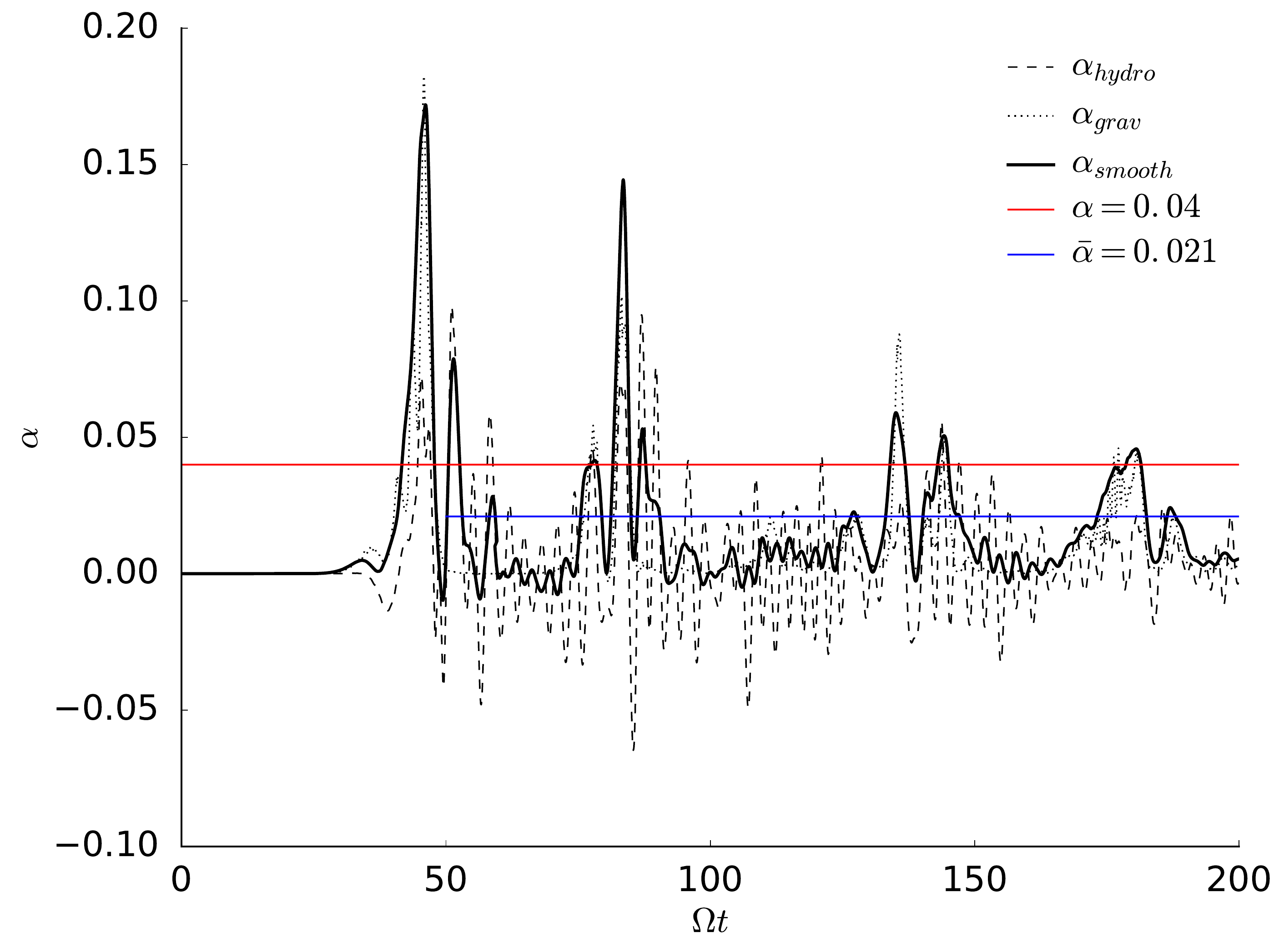}
\caption{The $\alpha$ stress and its gravitational and hydrodynamic components from our \textit{G256t10} simulation. In the gravitoturbulent steady state the stress is dominated by the volatile hydrodynamic component and significant deviations are a results of fragmentation and the dominance of gravitational stresses. The red solid line indicates the analytic expectation for $\alpha$ from Equation \eqref{eq:alphapara} and the blue line indicates the average $\alpha$ over the range the line spans}. For readability, the total $\alpha$ has been boxcar smoothed with a kernal of $40$.
\label{fig:alphacomp}
\end{figure}
\begin{figure}
\centering
\includegraphics[width=0.5\textwidth]{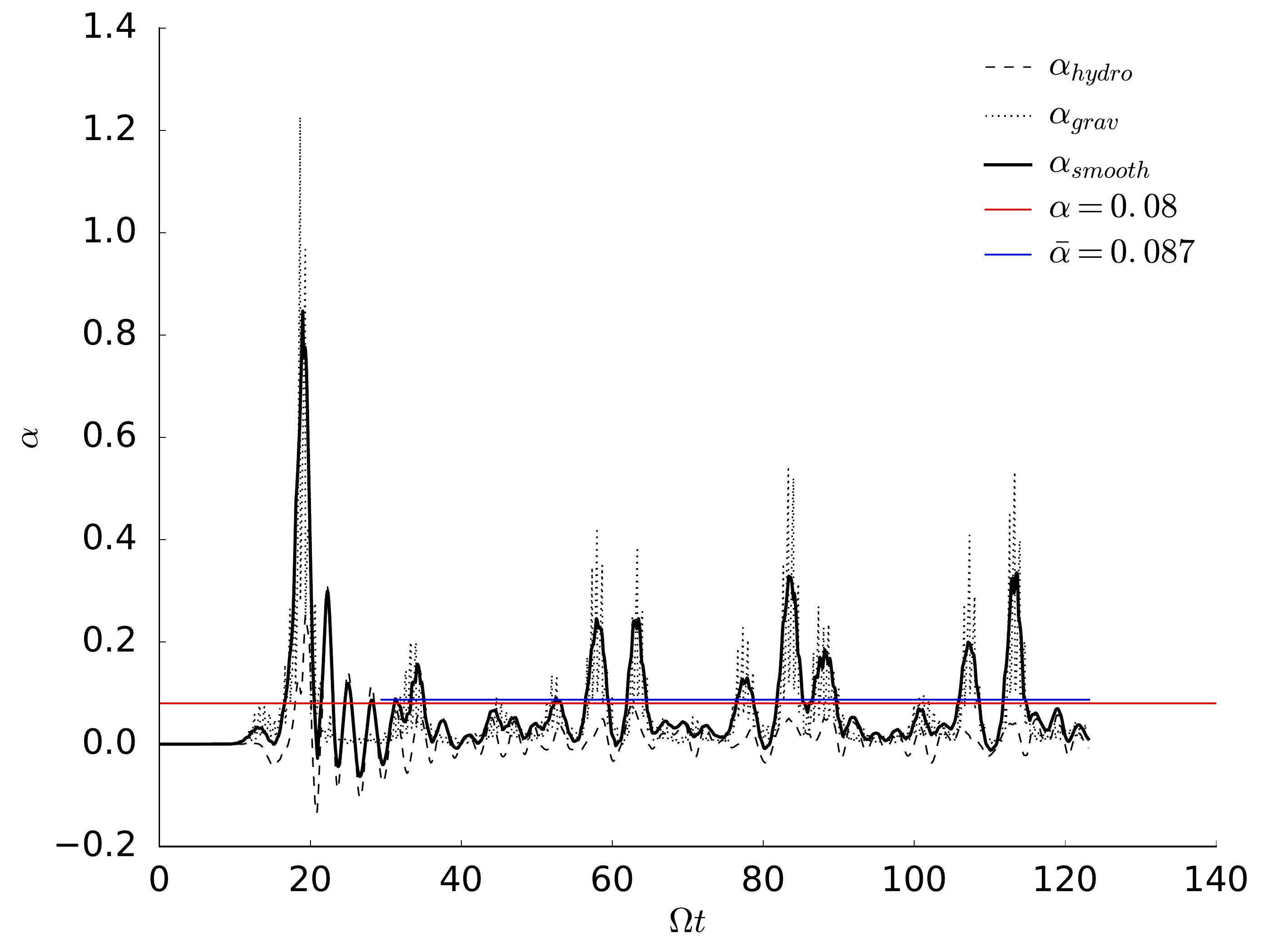}
\caption{The $\alpha$ stress and its gravitational and hydrodynamic components from our \textit{G512t5} simulation. Averaging the total stress over $t > 30 \Omega^{-1}$, when the simulation has settled, gives $\alpha=0.087$, within $10\%$.}
\label{fig:alphacomp512}
\end{figure}
 
\section{Results}
\label{sec:results}

As aforementioned, preliminary simulations of our 3D self-gravitating setup were conducted at our lowest resolution of $N^3 = 256^3$, which is the same effective resolution as the 2D simulations with $N^2=1024^2$ of \citet{Baehr2015}. When initialized with $Q_0 = 1$ and $\beta < 3$, these simulations show no gravitational instability, remaining largely static for tens to hundreds of orbital timescales. Self-gravity is not strong enough to excite gravitationally unstable modes. At our medium resolution of $N^3 = 512^3$, we found that $Q_0 = 1$ still shows no evidence for gravitational instabilities, suggesting that this is not a resolution dependent effect. For this reason, we use a new value of $Q_0 = 0.676$, a value close to what is typically assumed for disks of finite thickness \citep{Kim2002,Wang2010}.

We then conducted our simulations with this new $Q_0$ at three resolutions and at 4 cooling timescales, summarized in Table \ref{tab:sims}. The 3 different cooling timescales are selected assuming a critical fragmentation value of $\beta \sim 3$. For $\beta = 2$, we expect fragmentation in the classical scenario of 2D disk fragmentation \citep{Gammie2001} and for $\beta = 5,10,20$, we expect a stable gravitoturbulent disk, but are also concerned with the possibility of non-convergent effects, where fragmentation is possible for cooling timescales longer than $\beta = 3$.

Using this lower value of $Q_0 = 0.676$ immediately resulted in the unstable behavior expected from local disk instability simulations. While $Q_0=0.676$ is necessary to initiate GI due to the vertically stratified density (see Figure (\ref{fig:profileslQ})), it eventually settles to $Q=1$ as GI develops because the vertical structure of the disk becomes much less stratified as the disk warms up and $Q$ rises. This is shown in Figure (\ref{fig:toomre}), where the box-averaged Toomre stability parameter starts at the initial unstable $Q_0=0.676$ before stabilizing around $Q=1$. The initial oscillations correspond to the adjustment to hydrostatic equilibrium and do not affect the outcome. Figure \ref{fig:profileslQ} shows the average vertical density and temperature profiles over time in a fragmenting simulation at our medium resolution. Similar to the fiducial $Q=1$ simulations, the vertical density profile makes a slight adjustment to reach hydrostatic equilibrium and once gravitational instability sets in, mass concentrates in the midplane with time. 

Fragmentation was observed in our high resolution simulations for critical cooling timescales of  $2<\beta_{\mathrm{crit}} <5$. However, In contrast to \citet{Gammie2001}, our $N^3=256^3$ simulations never fragmented in either 2D or 3D, even at $\beta = 0.5$. That the high resolution cases agree with earlier results suggests the problem is with resolution rather than the introduction of the third dimension. We believe the lack of fragmentation in our low resolution runs might arise from higher numerical dissipation in Pencil as compared to ZEUS.

These low resolution runs also show that our models are not susceptible to premature or spurious fragmentation, an issue that affects simulations with poor resolution or inadequate error suppression. Spurious errors may produce density perturbations which may result in undesired fragmentation \citep{Truelove1997,Nelson2006} Inability to maintain a fragment for more than a few orbits does not necessary mean that we cannot say something about their ability to fragment however. As we will show in Section \ref{subsec:fragmentation}, one can derive a density threshold which designates the formation of a fragment, even if it is soon disrupted.

Figure \ref{fig:maxdensitylQ} shows the maximum density and average temperature for our $N^3=512^3$ (left) and $N^3=1024^3$ (right) simulations and demonstrates the anticipated fragmentation behavior of our 3D self-gravitating setups. At our medium resolution $N^3=512^3$ fragmentation occurs in the case where the cooling timescale is short enough to remove pressure support and the clump collapses before being torn apart by shear. For longer cooling timescales, the fragment cannot collapse fast enough to remain bound and is soon disrupted; the difference between the two states is illustrated in Figure \ref{fig:fragment}. Also shown in Figure \ref{fig:maxdensitylQ} are the box-averaged temperatures over time, showing that the fragmenting simulations are able to withstand the rising temperatures as the fragment collapses and becomes denser.

We ran identical initial conditions with at highest resolution, $N^3=1024^3$. The right plot of Figure \ref{fig:maxdensitylQ} shows the results of these simulations and they also show a fragmentation boundary of $\beta_{\mathrm{crit}} = 3$, showing that our models converge with resolution. 

Viscous stresses are only well parametrized in the case the disk is stable/gravitoturbulent, as is shown with the runs in Figures \ref{fig:alphacomp} and \ref{fig:alphacomp512}. Broken down into the hydrodynamic and gravitational components of the $\alpha$ stress, the total in Figure \ref{fig:alphacomp} averages to $\alpha = 0.21$ after settling at around $t = 50 \Omega^{-1}$. However, at the medium resolution in Figure \ref{fig:alphacomp512} the $\alpha$ stress agrees with Equation \eqref{eq:alphapara} to within $10\%$, suggesting that the inability of the $\alpha$ stress to reach expected levels at low resolutions is related to the inability to fragment and the higher numerical dissipation at lower resolution.

  \subsection{Fragmentation}
  \label{subsec:fragmentation}

In the 2D case, fragmentation can be easily established with a Roche surface density threshold (See section 2.4 of \citet{Baehr2015}), but this threshold does not translate to 3D simulations and a volume density threshold. Instead we find the pressure-modified Hill condition of \citet{Kratter2011} to result in a suitable threshold density. This threshold considers that pressure support stabilizes a fragment against collapse and leaves it more susceptible to tidal shear. Starting with the condition that a fragment has a radius that is a fraction of it's Hill radius
\begin{equation} \label{eq:fragmentradius}
r_{\mathrm{frag}} = \left( \frac{1}{f}\right)^{1/3} \left( \frac{M_{\mathrm{frag}}}{3M_{*}}\right)^{1/3} R,
\end{equation}
where $M_{*}$ is the mass of the central star, $R$ is the location away from the center and $f$ is factor determined by the choice of specific heat ratio $\gamma$. Defining the fragment mass as the integration of density over radial shells
\begin{equation} \label{eq:fragmentmass}
M_{\mathrm{frag}} = 4\pi \int_{h_{50}}^{r_{\mathrm{frag}}} \rho_{\mathrm{h}} \left( \frac{r_{\mathrm{h}}}{r}\right)^k r^2 dr,
\end{equation}
with the density at the center of the fragment $\rho_{\mathrm{h}}$ at a size one fiftieth the scale height $h_{50}$, a fragment outer radius of $r_{\mathrm{frag}}$ and a power law index of the fragment density distribution $k$. Then inserting the fragment mass into Equation \eqref{eq:fragmentradius} and assuming $h_{50} << r_{\mathrm{frag}}$, we then solve for the central density $\rho_{\mathrm{h}}$
\begin{equation} \label{eq:fragmentthreshold}
\rho_{\mathrm{h}} = \frac{3f(3-k)M_{*}}{4\pi R^3} \left( \frac{r_{\mathrm{frag}}}{h_{50}}\right)^k.
\end{equation}
We use $\gamma = 5/3$ which makes $f=4$, whereas a softer choice of $\gamma=7/5$ would mean $f=16$, resulting in a slightly higher central density. Equation \eqref{eq:fragmentthreshold} will serve as our threshold density for our simulations, provided that the maximum density is above this threshold for several timescales, thus separating fragments from overdensities which have not yet collapsed enough.

\begin{figure}
\centering
\includegraphics[width=0.5\textwidth]{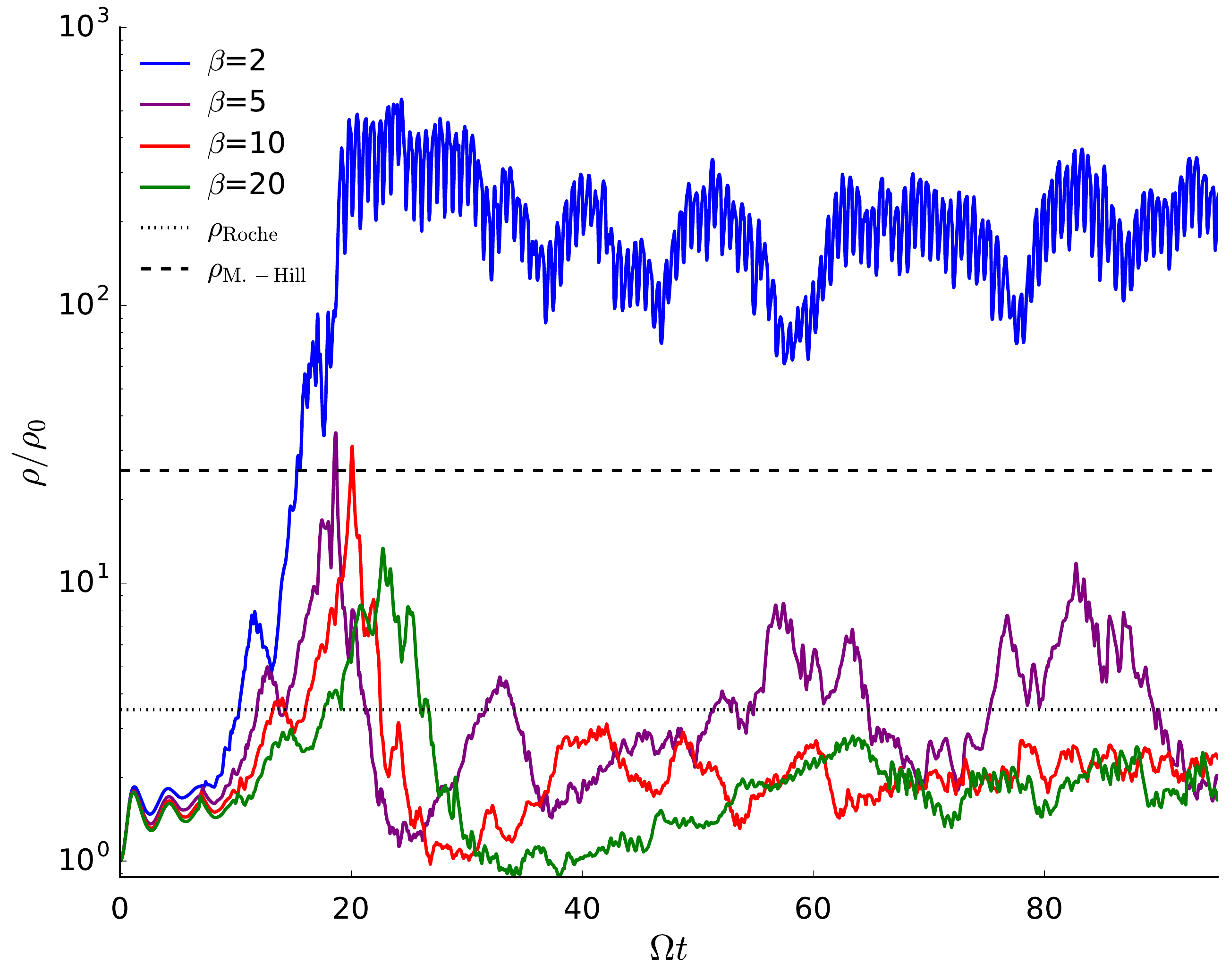}
\caption{Fragment threshold densities $\rho_{\mathrm{M.-Hill}}$ (dashed line) and $\rho_{\mathrm{Roche}}$ (dotted line) compared to our $N^3=512^3$ simulations for longer run times. The Roche threshold is clearly not an appropriate threshold in the non-fragmenting case, as the maximum density crosses this value only to fall below again later, whereas the pressure-modified Hill threshold is more consistent in delineating between gravitoturbulence and fragmentation.}
\label{fig:threshold}
\end{figure}

Finally, we wish to apply our fragment threshold, Equation \eqref{eq:fragmentthreshold}, to our simulations, from which we need to extract the density profile of a fragment. We find that a fragment scales with $k=0.7$ to a fragment radius of $r_{\mathrm{frag}} = 2/3 H$. With this, we find that for an overdensity in our simulations to withstand tidal disruption, as is shown in Figure (\ref{fig:threshold}), it must satisfy
\begin{equation} \label{eq:fragmentnumber}
\rho_{\mathrm{h}} \geq 25.57 \frac{M_{*}}{R^3},
\end{equation}
which, for a protoplanetary disk around a solar-mass star, results in a densities of $1.2 \times 10^{-10} \mathrm{g/cm^3}$ at $50 \mathrm{AU}$ and $1.5 \times 10^{-11} \mathrm{g/cm^3}$ at $100 \mathrm{AU}$. Comparing this to our simulations and to the Roche density in Figure \ref{fig:threshold} we see that this density threshold more adequately delineates between fragments and non-fragments. A disk with the temperatures as in these simulations will have an aspect ratio $H/R \approx 0.11$ at $50 \mathrm{AU}$ and $H/R \approx 0.16$ at $100 \mathrm{AU}$. With these densities and aspect ratios, we approximate threshold fragment masses of $5 M_{\mathrm{J}}$ and $15 M_{\mathrm{J}}$ at $50$ and $100$ AU, respectively.

\begin{figure}[h]
\centering
\includegraphics[width=0.5\textwidth]{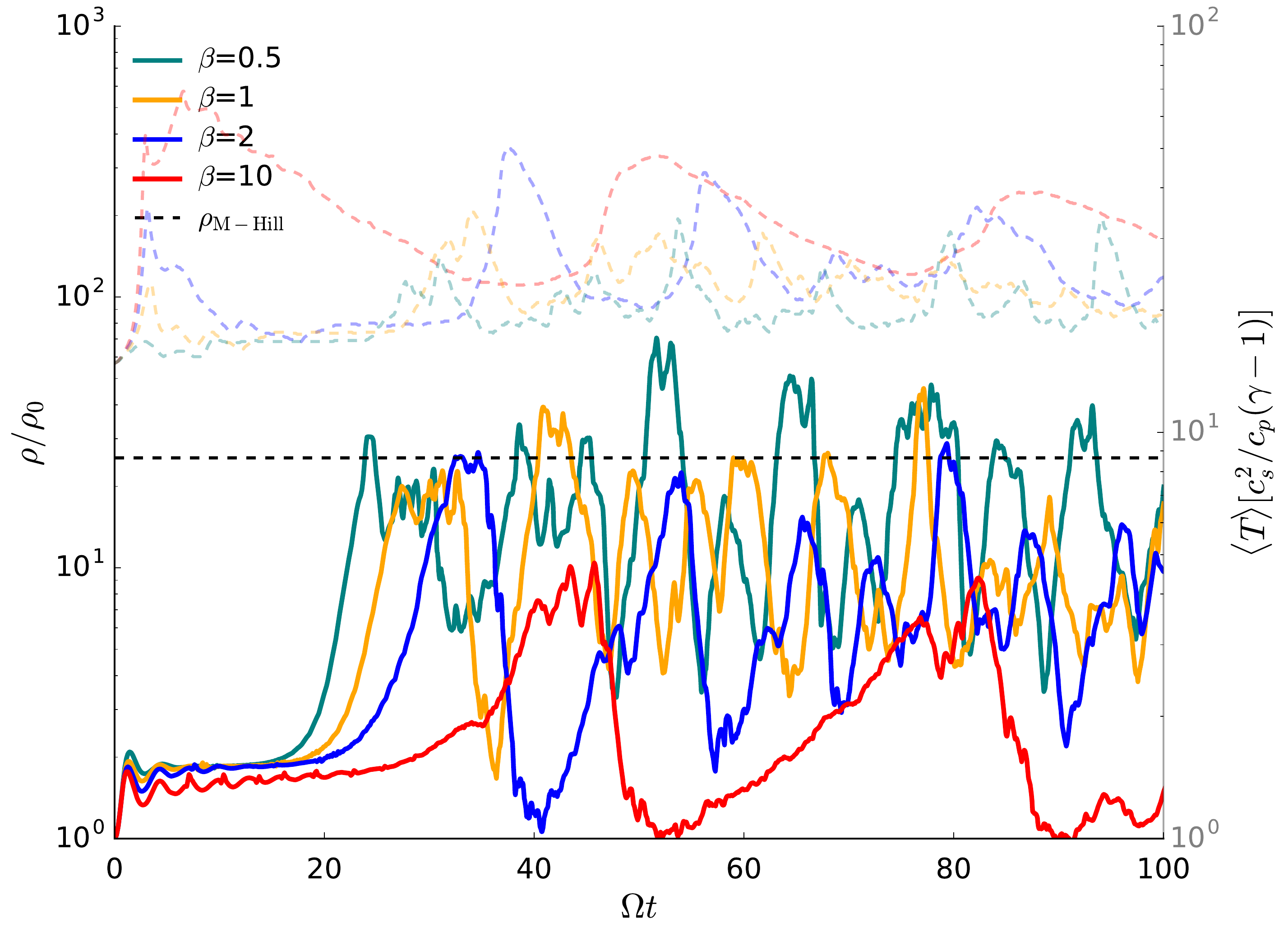}%
\caption{Time series of the maximum surface density in a set of simulations carried out at resolution $N^3=256^3$ and $Q_0 = 0.676$. Faint dashed lines are the box-averaged temperatures corresponding to the appropriate density curve by color. The black dashed line is the pressure modified Hill density criterion defined by Equation \eqref{eq:fragmentnumber}.}
\label{fig:density256lQ}
\end{figure}

Applying the density threshold to our low resolution simulations which failed to form sustained fragments, as one can see in Figure \ref{fig:density256lQ}, simulations with very short cooling times ($\beta =1,0.5$) surpass this threshold. If the maximum density of a fragment can surpass this critical density for more than an orbit, it can be considered a fragment, even if it is subsequently disrupted. Thus, this threshold may provide a useful sink cell formation criterion for simulations with resolution too low to fragment.

\section{Discussion}
\label{sec:discussion}
\begin{figure*}[t]
\centering
\includegraphics[width=0.5\textwidth]{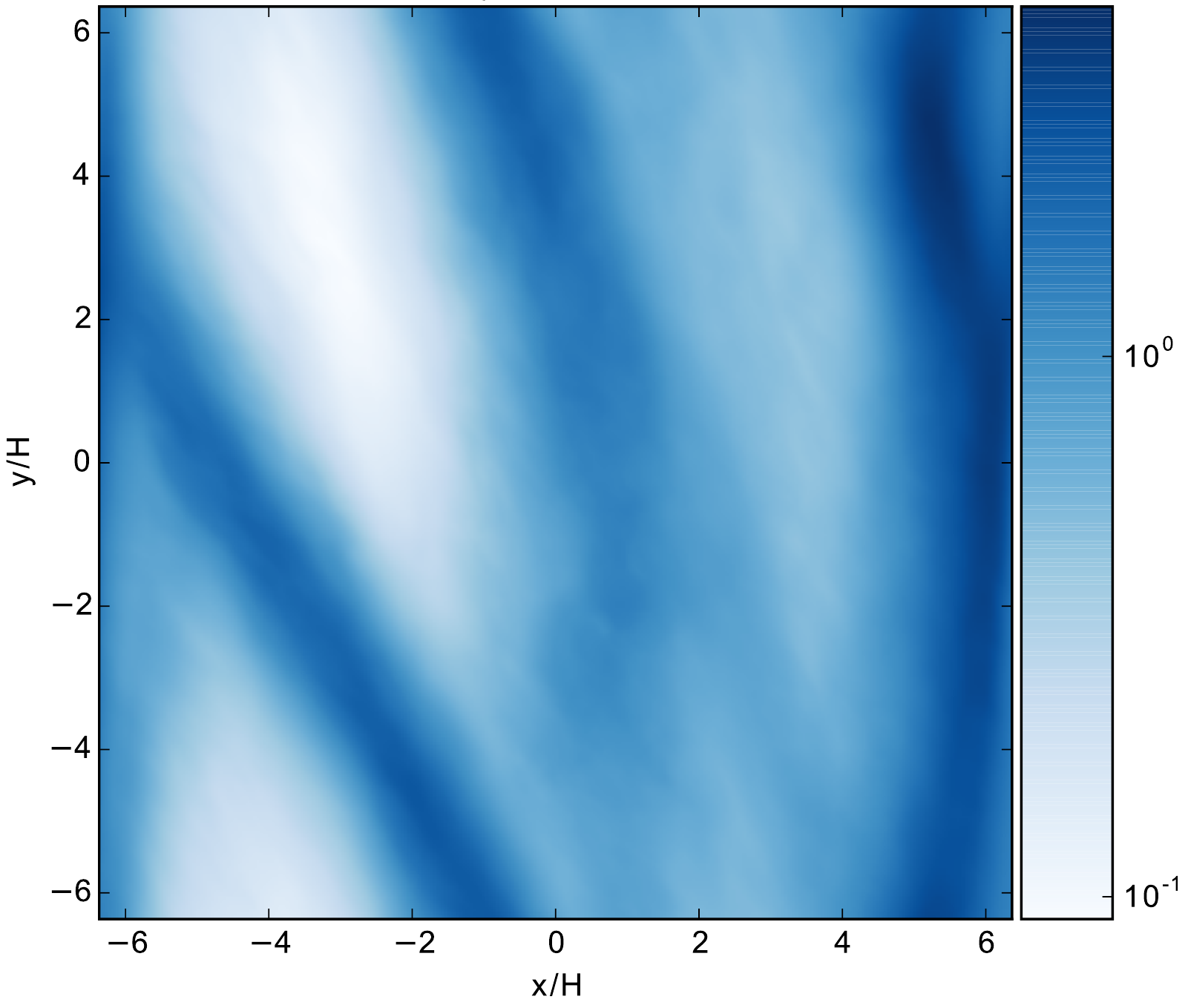}%
\includegraphics[width=0.5\textwidth]{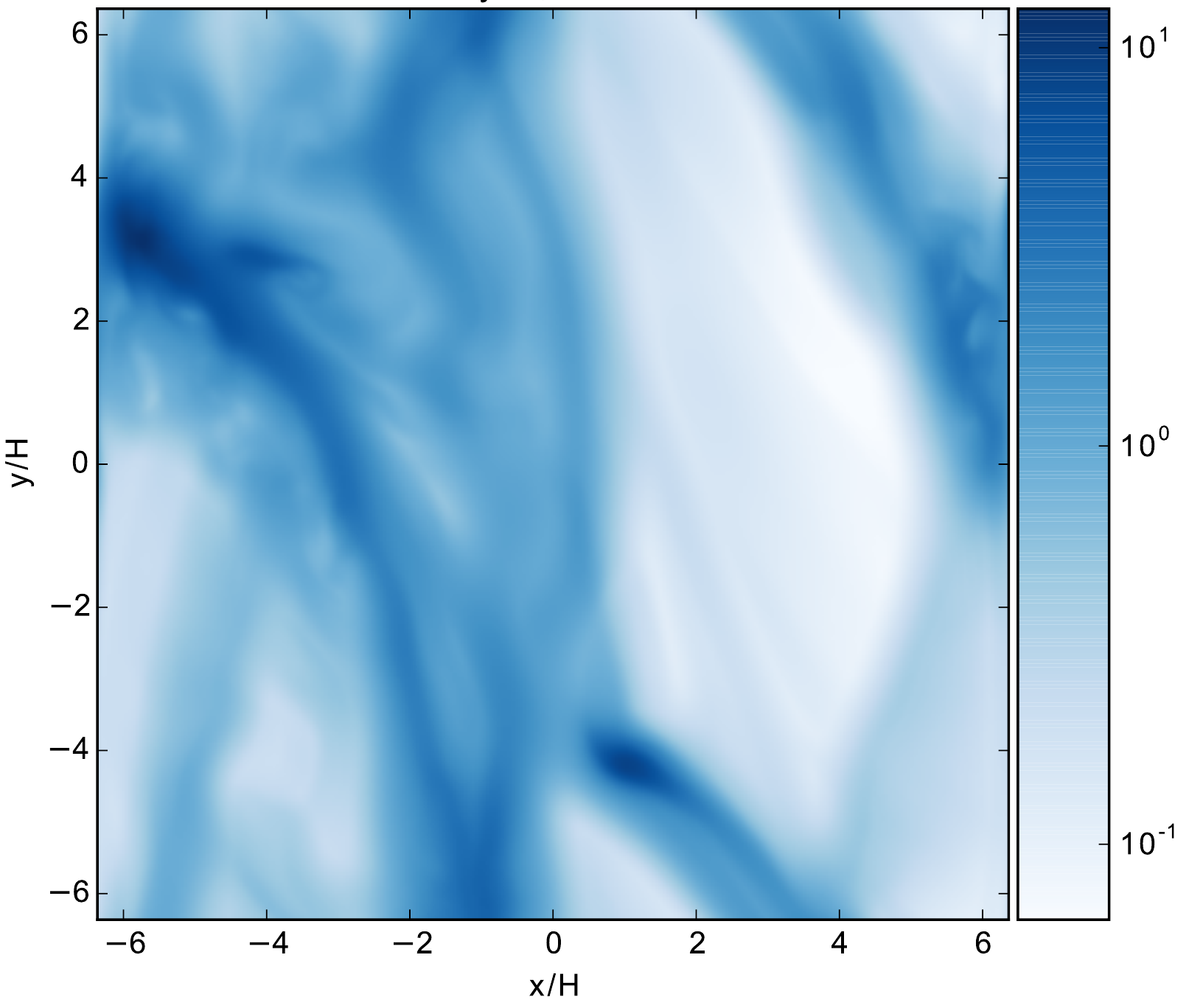}%
\caption{\textit{Left}: Map of surface density from a low resolution ($N^3=256^3$) 3D run with efficient cooling $\beta=2$. With such a short cooling timescale, this setup is expected to fragment but instead shows sheared self-gravitating filaments typical of gravitoturbulence. \textit{Right}: Map of surface density from a 2D simulation, which similar to the 3D case fails to fragment but is slightly clumpier due to the stronger self-gravity. The simulation shown on the right is the same as the models from \citet{Baehr2015}, but scaled down to mimic the size and resolution of the 3D runs presented in this paper.}
\label{fig:density3d}
\end{figure*}
  \subsection{Convergence}
  \label{subsec:converge}
  
Following the results of \citet{Young2015}, which demonstrated the need for a smoothing length to achieve convergent disk fragmentation in thin disk simulations, we expected our 3D simulations would converge since a smoothing length is no longer required for a disk with finite thickness. There might be additional effects which affect fragmentation at different resolutions, including but not limited to cooling, viscosity and numerical effects, but investigating the extension of local self-gravitating disks to include a vertical component is the most natural extension of that work and our results support their findings in this regard.

Recent work has focused on numerical effects and their influence on non-convergence. \citet{Klee2017} found that numerical oversteepening could be responsible for the overestimation of overdense peaks resulting in fragmentation. Through our use of sixth-order hyperdiffusion, we do not expect oversteepening to be a significant factor. \citet{Deng} find that dissipation of angular momentum in SPH methods removes shear support from regions with high flow velocities and as this depends on resolution, fragmentation becomes easier with increasing resolution.
  
While our results strongly suggest convergence, at our lowest resolution fragmentation is suppressed due to insufficient resolution for both 2D and 3D models, similar to the results of \citet{Turk2012}, which found that turbulent velocity fluctuations suppressed the formation of small scale magnetic dynamos when the Jeans length is not resolved by at least 64 grid cells. At higher resolutions the results follow the standard fragmentation scenario, with clear fragmentation at $\beta = 2$ and a lack of fragmentation evident at $\beta = 5,10,20$. The $\beta$ cooling prescription is not a true substitute for realistic cooling, only a simple parametrization of the underlying physics. The use of a fixed background term may have an effect on convergence \citep{Rice2012a}, thus one must be careful that convergence is not lost if this term is removed \citep{Lin2016}. However, since simulations using $\beta$ cooling and background irradiation converge, it supports our understanding of the physics.

  \subsection{Reconciling with the Thin Disk Approximation}
  \label{subsec:3dvs2d}
  
A large part of this investigation has been about the differences between 2D and 3D simulations of self-gravitating disks, but it is also necessary to discuss how they are similar. Vertical structure and the accompanying gravitational dilution mean disk instability and fragmentation require even higher gas surface densities to collect locally before GI can set in. Once instabilities develop, the results of 2D simulations with smoothed self-gravity and 3D simulations show similar structures and features on similar scales (Figure (\ref{fig:density3d})). Gravitoturbulent simulations in 3D show non-axisymmetric structures of the same scale as in 2D, but do not show as strong a contrast as 2D simulations due to the weaker self-gravity on scales smaller than $H$.

Ultimately, both 2D and 3D simulations converge to the same critical cooling timescale but 3D simulations take significantly more resources. While 3D simulations are convergent and can capture effects at small scales, there is growing evidence that scales smaller than $\sim H$ contribute little to the growth of gravitationally unstable modes \citep{Young2015,Riols}. Thus, for the modeling of self-gravitating disks, 2D simulations may be sufficient so long as an appropriate smoothing length is used to account for the vertical scale of the disk.

  \subsection{Fragmentation and Planet Formation}
  \label{subsec:form}
  
Planet formation by gravitational instability is hindered by the large disk masses required. With the requisite $Q_0$ for fragmentation decreasing in these simulations, it is be even more difficult to form planets by GI by requiring either more mass or a colder disk for fragmentation. Larger relative disk sizes required for fragmentation also suggest that the resulting fragments will draw from a larger mass reservoir to overcome stabilizing effects and form larger objects. Thus, it is more likely fragments already have enough mass to be considered brown dwarfs or low-mass stars, without taking into account post formation accretion and growth. 

The density threshold of Equation \eqref{eq:fragmentnumber} and the subsequent fragment mass suggest objects formed in these simulations are formed with at least $4.9 \mathrm{M_J}$ at both $50 \mathrm{AU}$ and $100 \mathrm{AU}$ around a solar-like star. Since companions will generally form larger than these thresholds and subsequently accrete more mass, these fragments will likely form planets many Jupiter masses in size and brown dwarfs as well. Recent observations show fragments are indeed quite large, as is the case of L1448 IRS3B \citep{Tobin2016} which is consistent with theoretical expectations \citep{Stamatellos2009a}.

  \subsection{Further Considerations}
  \label{subsec:further}
  
Global simulations can model the growth, migration, and potential destruction of fragments formed in a disk, but with our shearing box simulations are able to resolve the small scales of GI structure and development. For a resolution and convergence study this makes local simulations ideal for the relatively small patches of the disk where fragmentation occurs.

The Fourier solver in Pencil is limited to domains with the same number of grid cells in the vertical direction as in the \textit{x}-direction, meaning that a simulation with a flattened box would have significantly higher resolution in \textit{z}, resulting in different values of hyperviscosity in each direction, possibly leading to unphysical effects. For that reason, we conducted our simulations spanning roughly $12.7$ scale heights in the vertical direction, much of which was not crucial to the simulation of fragmentation. Accommodating these largely empty spaces require carefully establishing a minimum density low enough to prevent undesired additional mass from settling onto the disk and affecting fragmentation. A more efficient approach would be to simulate the vertical direction in $4$ to $6$ scale heights while still retaining the radial and azimuthal extents. Additionally, the periodic vertical boundary conditions for the self-gravitational potential rather than typical outflow conditions and may contribute to the somewhat erratic $\alpha$ stresses calculated in Figure \ref{fig:alphacomp512}.

While the simple cooling with irradiation used here might result in convergence of the cooling criteria, a better description of collapse and fragmentation will ultimately come from using radiative transfer calculations, whether in the form of approximations such as flux-limited diffusion \citep{Mayer2007,Boley2006} or more sophisticated but intensive methods like ray tracing. These methods take into account the gas opacities and optical depths to more appropriately model variations of cooling from one region of the disk to the next.

One feature of protoplanetary disks which is typically ignored in self-gravitating contexts but might have an effect on fragmentation is the disk magnetic field. While ionization is expected to be low in the cold outer regions of circumstellar disks, the effect of magnetic fields on fragmentation is uncertain. Only a pair of studies have looked into the affects of magnetic fields on the fragmentation criterion \citep{Riols2016,Forgan2017}, and came to opposite conclusions about the importance of magnetic fields on disk fragmentation.

\section{Conclusions}
\label{sec:conclusion}

In this paper we have conducted a converge study of 3D shearing box self-gravitating disks with simple cooling and background irradiation. Starting with a marginally unstable disk, simulations were run at various cooling timescales through the initial burst phase where they have either formed a fragment or settled to gravitoturbulence. These simulations were run at several high resolutions to investigate fragmentation conditions and the convergence of the solution with respect to the cooling timescale and our results are as follows:

\begin{enumerate}
\item Consistent with analytic results, we find that 3D vertically stratified disk simulations fragment for two high resolution simulations with a cooling timescale boundary of roughly $\beta_{\mathrm{crit}} = 3$. Fragmentation only occurs when a scale height is resolved by at least 40 grid cells.
\item Convergence indicates that there is a firm criterion for fragmentation in a disk: where disks cool rapidly with $\beta < 3$. This limits fragmentation to the distant regions around a star where temperatures are relatively low and dominated by stellar irradiation.
\item By achieving convergence, we have demonstrated that the standard $\beta$ cooling model with a fixed floor temperature may have no inherent flaws numerically when used in our 3D models. It is nevertheless desirable to use a more realistic model for disk cooling in future studies.
\item Using a pressure-supported Hill criterion, we find a suitable minimum density consistent with the formation of brown dwarfs and low-mass stars. This serves as a useful criterion for the inclusion of sink cells to aid in the efficient computation of fragmentation as it can be applied to simulations with low resolution simulations which are unable to fragment.
\end{enumerate}

\begin{acknowledgments}

The authors would like to thank Chao-Chin Yang and Andreas Schreiber for useful discussions and assistance with the Pencil code and the anonymous referee for comments which improved the paper. KMK received support for this research from the Max Planck Institute for Astronomy. Simulations shown were run on the Theo/Isaac clusters at the Rechenzentrum Garching (RZG) of the Max Planck Society and the JUQUEEN cluster of the J{\"u}lich Supercomputing Centre \citep{Stephan2015}.

\end{acknowledgments}

\bibliography{library}

\end{document}